# Femtosecond intramolecular rearrangement of the CH$_3$NCS radical cation

Jacob Stamm[1], Shuai Li[1], Bethany Jochim[1], Stephen H. Yuwono[1], Swati S. Priyadarsini[1], Piotr Piecuch[1,2], and Marcos Dantus[1,2,*]

[1] Department of Chemistry, Michigan State University, East Lansing, MI 48824, USA
[2] Department of Physics and Astronomy, Michigan State University, East Lansing, MI 48824, USA
* Corresponding author: dantus@msu.edu

**Abstract**

Strong-field ionization, involving tunnel ionization and electron rescattering, enables femtosecond time-resolved dynamics measurements of chemical reactions involving radical cations. Here, we compare the formation of CH$_3$S$^+$ following the strong-field ionization of the isomers CH$_3$SCN and CH$_3$NCS. The former involves the release of neutral CN, while the latter involves an intramolecular rearrangement. We find the intramolecular rearrangement takes place on the single picosecond timescale and exhibits vibrational coherence. Density functional theory and coupled-cluster calculations on the neutral and singly ionized species help us determine the driving force responsible for intramolecular rearrangement in CH$_3$NCS. Our findings illustrate the complexity that accompanies radical cation chemistry following electron ionization and demonstrate a useful tool for understanding the cation dynamics after ionization.

## I. Introduction

Methyl isothiocyanate (MITC, CH$_3$NCS) is the primary decomposition product of metam potassium (CH$_3$NHCS$_2$K). It is the most widely used agricultural herbicide, insecticide, fungicide, and nematicide, with worldwide use approaching $10^6$ metric tons per year in 2019.[1] Its vapor is quickly decomposed by sunlight to produce CH$_3$NC + S with almost unity quantum yield.[2] Methyl thiocyanate (MTC, CH$_3$SCN), on the other hand, is an extremely hazardous compound not used in agriculture. Studies following photodissociation of these two compounds at 193 nm and 248 nm suggest that both share a common excited electronic state which can produce CH$_3$S + CN.[3] In this work, we focus on the fragmentation of MITC and MTC following strong-field ionization to determine if a shared potential energy surface in the ionized state is the reason why the electron ionization mass spectra of these two distinct compounds are so similar.

During strong-field ionization,[4–6] the laser field pulls the most labile electron away from the source molecule. When the electric field reverses direction, the ejected electron is accelerated toward the originating atom or molecule with tens of eV of energy. This event is more likely to occur within the same optical cycle (2.67 fs for 800 nm photons) than in subsequent cycles. While first demonstrated in smaller systems, decades of studies on strong-field rescattering and high-harmonic generation (HHG) have shown that these phenomena is commonplace during strong-field processes involving polyatomic molecules.[7–15] Femtosecond pulses of 800 nm central wavelength and $1\times10^{14}$ W cm$^{-2}$ peak intensity can tunnel ionize large polyatomic molecules with ionization potentials (IPs) ranging from 8 to 10 eV, creating high-energy electrons that can deposit much of their energy back into the molecule upon rescattering. The maximum kinetic energy of the rescattering electrons is proportional to the laser intensity, approximately 3.2 $U_p$,[16] where $U_p$ is the ponderomotive energy. For the above laser parameters, this maximum energy is about 19 eV.

In this study, the dynamics observed following ultrafast strong-field ionization may shed light on the fragmentation processes occurring in electron ionization mass spectrometry (EI-MS). This is because the ultrafast excitation driven by the 70 eV EI-MS electrons results in a broad distribution of internal energies of the molecule, varying, depending on the molecular size, between 10 and 50 eV due to the wide range of impact parameters.[17,18] This range of internal energies quickly leads to single or multiple ionization followed by fragmentation and intramolecular vibrational energy redistribution (IVR) occurring on fs-to-ns timescales. Similarly, electron rescattering leads to a broad range of internal energies. Provided we normalize to the amount of energy deposited into the molecule by matching the fragmentation pattern observed in EI-MS, electron rescattering leads to ionization similar to that of EI-MS. However, in the case of femtosecond ionizing pulses, the ionization occurs on the timescale of ~10 fs, thus enabling femtosecond time-resolved studies relevant to the fragmentation mechanisms occurring in EI-MS. In this study, we focus on the time-resolved rearrangement reaction dynamics in CH$_3$NCS following strong-field ionization. Our experimental work is augmented by quantum chemistry calculations for the CH$_n$NCS$^+$, $n = 0$–3, species to search for possible rearrangement mechanisms of the CH$_n$NCS$^+$ reactants prior to the formation of the corresponding CH$_n$S$^+$ product ions.

**II. Experimental and Computational Details**

The experimental apparatus employed in this study has been described in a previous publication.[19] Briefly, a Ti:Sapphire (Coherent, Legend) 40 fs laser with a 1 kHz repetition rate and 800 nm central wavelength, is split into pump and probe pulses by a 80:20 beam splitter. The temporal delay between the pump and probe pulses was controlled by a translation stage (Aerotech, ANT130L). The pump and probe pulses were recombined and focused into the interaction region by an $f$ = 300 mm achromatic lens. The pump intensity was selected to optimize ion yields without saturating the larger mass-to-charge fragments. The peak intensity of the pump was $1\times10^{14}$ W cm$^{-2}$ and calibration of the laser intensity was performed using the $N_2^{2+}/N_2^+$ and $Ar^{2+}/Ar^+$ ion yield ratios.[20,21] The wavelength, intensity, and pulse duration of the pump corresponded to two values of Keldysh parameter $\gamma$ of 0.88 and 0.92 (assuming an IP of 9.25 and 10 eV of MITC and MTC, respectively[22]), which favor tunnel ionization.[23] In this intermediate ionization regime, some combination of tunnel and multiphoton ionization may occur.[24] However, field ionization is different for a multi-well system such as molecule, compared to the single Coulomb potential well of an atom. In particular, molecular ionization is highly dependent on interatomic distances.[25] Furthermore, combined theoretical and experimental studies have found that tunnel ionization in molecules occurs at lower laser intensities than for atoms with similar IP.[26,27] For example, in a study on acetone, butyl-acetone, and 3-pentanone, it was found that tunnel ionization takes place at ~$6\times10^{13}$ W cm$^{-2}$ when using 800 nm femtosecond pulses of similar duration as used in our study.[28] We have also observed similar dynamics to those reported in the present work when using higher pump intensities ranging from $2\times10^{14}$ to $6\times10^{14}$ W cm$^{-2}$ (i.e., $\gamma$ of 0.62 going down to 0.36).

The dynamics of all the different products following ultrafast ionization were tracked using disruptive probing with the weak 800 nm pulse.[29] The probe pulse was polarized at the "magic" angle (54.7 degrees) relative to the pump pulse to minimize the influence of rotational dynamics on the measurements. The probe was attenuated to about $3\times10^{13}$ W cm$^{-2}$ to ensure that it did not generate ions on its own. This way, the probe pulse can only disrupt the chemical reaction of interest while the chemical transformation is happening, preventing or altering its completion. The ability to deplete the product of interest could thus be tracked with femtosecond time resolution, yielding information about the dynamics involved. Once the product was formed, it could no longer be depleted by the weak probe pulse. In our experiments, the pump pulse caused ionization and fragmentation, and the weak probe pulse could only disrupt the product yield if it arrived before such a product was completely formed. Therefore, disruptive probing provided information about the timescale of product formation.

The samples of $CH_3NCS$ and $CH_3SCN$ underwent several freeze-pump-thaw cycles before the sample vapor was introduced into the chamber through a needle valve as an effusive beam. During all the measurements, the pressure of the mass spectrometer was kept below $5\times10^{-6}$ torr. When the sample needle valve was closed, the background pressure dropped quickly to the $10^{-8}$ torr range, whereas the base pressure was in the $10^{-9}$ torr range. The measurements were performed using a home-built time-of-flight (TOF) mass spectrometer,[30] in which ions were detected using a Chevron-configuration microchannel plate detector. The ion signals were digitized by an oscilloscope (LeCroy, WaveRunner 610Zi). A 1 mm slit in the extractor plate was used to limit the ion collection region and mitigate focal-volume averaging effects, given a Rayleigh length of about 1.8 mm. For every time delay in one time-resolved scan, a TOF spectrum was obtained by averaging over 1,000 laser shots. Each time-resolved plot is the average of several hundred iterations of a time-resolved scan, and thus every data point is an average of more than 75,000 laser shots.

To assist the analysis of the experimental results, we performed quantum chemistry computations aimed at investigating possible intramolecular rearrangement pathways of the $CH_nNCS^+$ ions with $n = 0$–3, resulting from the ionization of $CH_3NCS$, that might potentially lead to the formation of the $CH_nS^+$ + CN products. We started by performing geometry optimizations and determination of harmonic vibrational frequencies for the parent $CH_3NCS$ molecule and the resulting $CH_nNCS^+$ ($n = 0$–3) ions in their respective ground electronic states using the Kohn–Sham formulation[31] of density functional theory (DFT)[32] employing the B3LYP[33,34] functional. We then searched for normal mode or modes that could lead to the migration of the $CH_n$ moiety from N to S in each of the $CH_nNCS^+$ ions and utilized the intrinsic reaction coordinate (IRC) approach to find the reaction pathways that connect these ions with the corresponding $[CH_nS\cdots CN]^+$ intermediate product species, in which the $CH_n$ group, originally attached to nitrogen, is transferred to the sulfur atom. To estimate the energetics of the final state associated with each reaction pathway, we also optimized the geometries of the $CH_nS^+$ ($n = 0$–3) and CN products using the B3LYP functional. To examine the effect of the electron correlation treatment on our results, the electronic energies at the stationary points along each reaction pathway obtained with B3LYP were recalculated using a high-level coupled-cluster (CC)[35] approach with singles, doubles, and noniterative triples defining the CR-CC(2,3) method of Refs. 36 and 37. To further enhance our discussion, we also computed vertical ionization energies corresponding to higher excited states of the $CH_3NCS^+$ ion at the $CH_3NCS$ geometry using the IP-EOMCC(3h-2p) approach developed in Refs. 38 and 39, which belongs to a larger category of ionization potential equation-of-motion CC methods.[40]

All electronic structure calculations reported in this work, which were performed using the GAMESS software package,[41,42] employed the cc-pVTZ basis set[43,44] with an additional tight d function for the S atom.[45] In the B3LYP calculations of the reaction pathways, we used the restricted and restricted open-shell formulations of Kohn–Sham DFT. The CR-CC(2,3) calculations using the restricted Hartree–Fock (for all singlet ions and molecules) and restricted open-shell Hartree–Fock (for all non-singlet species) determinants as reference functions, along with the IP-EOMCC(3h-2p) computations of the lowest few ionization energies of the $CH_3NCS$ molecule, were performed using the routines developed by the Piecuch group,[36–39,46] which form part of the GAMESS code. In all CC computations, the core orbitals corresponding to the 1s shells of the C and N atoms and the 1s, 2s, and 2p shells of the S atom were kept frozen. The IRC calculations employed the Gonzalez–Schlegel second-order method,[47] which is the default option in GAMESS. The Cartesian coordinates for the stationary points along all reaction pathways determined in this work are provided in the Supplementary Material.

### III. Results and Discussion

The mass spectra for $CH_3NCS$ and $CH_3SCN$ following EI-MS and strong-field ionization are shown in Fig. 1. We find that the EI-MS spectra for both species are quite similar, albeit with a reduced yield of $CH_nS^+$ ($n = 0–3$) ions in the case of $CH_3NCS$, a difference that is reflected in the strong-field ionization spectra as well. The strong-field ionization spectrum of MITC ≡ $CH_3NCS$ shows a prominent peak at 35.5 m/z, corresponding to the doubly ionized MITC and loss of $H_2$, a process with an appearance energy of 28 eV.[48] We find that upon ionization MITC isomerizes and produces the $CH_nS^+$ ions with $n = 0–3$. We have confirmed that these products originate from MITC and not from contamination by its MTC isomer, $CH_3SCN$, as discussed below.

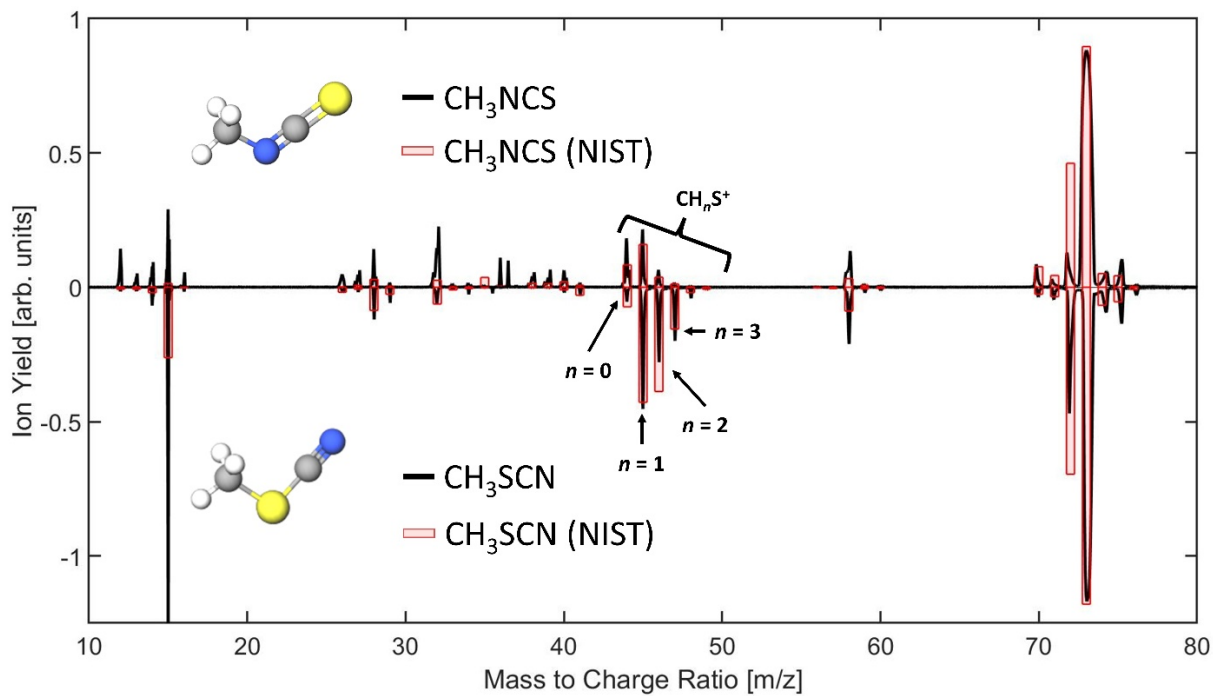

**Fig. 1**. EI-MS from the NIST database (red) and strong-field ionization (black) spectra for $CH_3NCS$ (top) and $CH_3SCN$ (bottom). The $CH_nS^+$ ($n = 0$–$3$) products of the rearrangement reaction are indicated by the black curly bracket.

A significant fraction of the electron energy is imparted to the parent ion through rescattering. The appearance of fragment ions in the mass spectrum provides an indication of the energy transfer from the returning electrons and the molecular ion's internal energy distribution after rescattering. Table I lists the appearance energy for multiple product ions following the electron ionization of MITC and MTC. Based on the presence of these fragments in our measurements, we can estimate the internal energy of the molecules following electron rescattering to be greater than 15 eV, and as high as 28 eV for MITC, given that we observe the doubly charged $CHNCS^{2+}$ species.

**Table I.** Appearance energies for $CH_3NCS$ and $CH_3SCN$ and ionization energies of some expected fragments.

**Appearance Potentials in eV**

**$CH_3NCS$**

| | |
|---|---|
| → $CH_3NCS^+$ | 9.25 ± 0.03 Ref. 23 |
| → $CH_2NCS^+$ | 11.9 ± 0.2 Ref. 48 |
| → $CNCS^+$ | 14.1 ± 0.3 Ref. 48 |
| → $NCS^+$ | 14.9 ± 0.5 Ref. 48 |
| → $CH_3^+$ | 15.3 ± 0.3 Ref. 48 |
| → $CS^+$ | 15.6 ± 0.15 Ref. 48 |
| → $CHNCS^{2+}$ | 28.0 ± 0.5 Ref. 48 |

**$CH_3SCN$**

| | |
|---|---|
| → $CH_3SCN^+$ | 9.96 ± 0.05 Ref. 51 |
| → $CH_2SCN^+$ | 12.6 ± 0.1 Ref. 52 |

**Fragment Ionization Energies in eV**

| | |
|---|---|
| $CH_3 \to CH_3^+$ | 9.84 Ref. 53 |
| $SCN \to SCN^+$ | 10.69 Ref. 54 |
| $CH_2 \to CH_2^+$ | 10.40 Ref. 53 |
| $CN \to CN^+$ | 14.17 Ref. 55 |
| $H_2 \to H_2^+$ | 15.43 Ref. 56 |
| $H \to H^+$ | 13.60 Ref. 53 |

We can determine if the isomerization reaction takes place following single or double ionization based on intensity difference spectra (IDS) of the different product ions.[49] The ion yields for MITC obtained at different intensities for three different ions are shown in Fig. 2 (a). By IDS, we mean that we quantify the finite differences $\Delta Y_i(I_H) = Y_i(I_H) - Y_i(I_L)$, where $Y_i(I_H)$ and $Y_i(I_L)$ are the yields of a particular ion of interest at higher and lower peak intensities, respectively (Fig. 2 (b)). The IDS spectra mitigate contributions from focal volume averaging by narrowing the range of intensities that drive a specific process identified in a spectrum. We find that the rearrangement reaction resulting in $CH_3S^+$ coincides with single ionization of the molecular ion and not with the doubly ionized species. A reference ion that is known to be produced from double ionization ($CH_3NCS^{2+}$) has distinctly different IDS spectra, peaking at $\sim 3 \times 10^{14}$ W cm$^{-2}$. The $CH_3S^+$ ion coincides with the singly ionized species $CH_3NCS^+$, peaking at $\sim 1 \times 10^{14}$ W cm$^{-2}$.

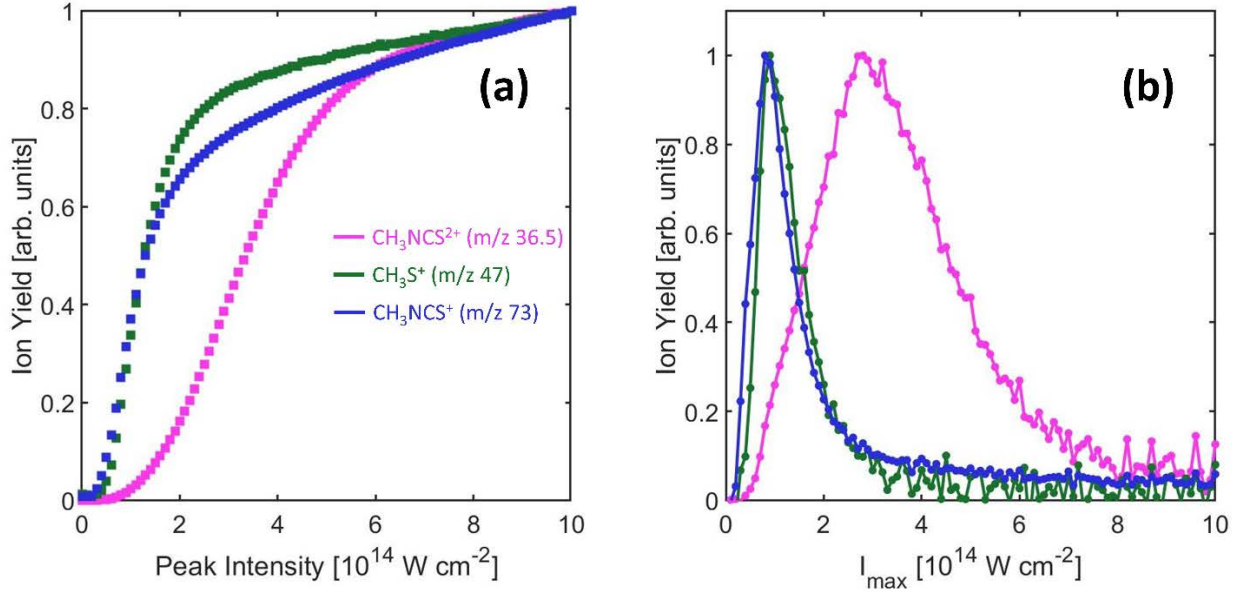

**Fig. 2**. (a) Yields of several key ions as a function of peak laser intensity and (b) IDS spectra of several key ions, which were obtained by subtracting each data point in (a) from the previous data point, effectively giving the slope of the ion yield vs peak intensity curve.

The time-resolved ion yields of $CH_3S^+$ following the strong-field ionization of both isomers under investigation are shown in Fig. 3. Each of the two traces is fitted to the following function:[29]

$$P(t, \tau_1, \tau_2, \tau_3) = P_1(t, \tau_1) + P_2(t, \tau_2) + P_3(t, \tau_3), \quad (1)$$

where

$$P_i(t, \tau_i) = A_i e^{-\frac{t}{\tau_i}} \left(1 + erf\left(\frac{t}{s} - \frac{s}{2\tau_i}\right)\right). \quad (2)$$

In the above equations, $t$ indicates a specific pump-probe delay, $\tau_i$ is a time constant associated with a rise ($\tau_{rise}$) or decay ($\tau_{decay}$) of the signal, $A_i$ is an amplitude factor, and $s$ is a parameter associated with the pump pulse duration through the expression $\tau_{FWHM} = 2s\sqrt{\ln 2}$. Equation (2) is the solution of the differential equation that describes the population of the ion state produced by the excitation with a Gaussian pulse with duration $\tau_{FWHM}$, whose decay is characterized by the time constant $\tau_{decay}$. When oscillations are present, the $P(t)$ function is subtracted from the data and fit to a decaying cosine function with the appropriate amplitude, frequency, and phase.

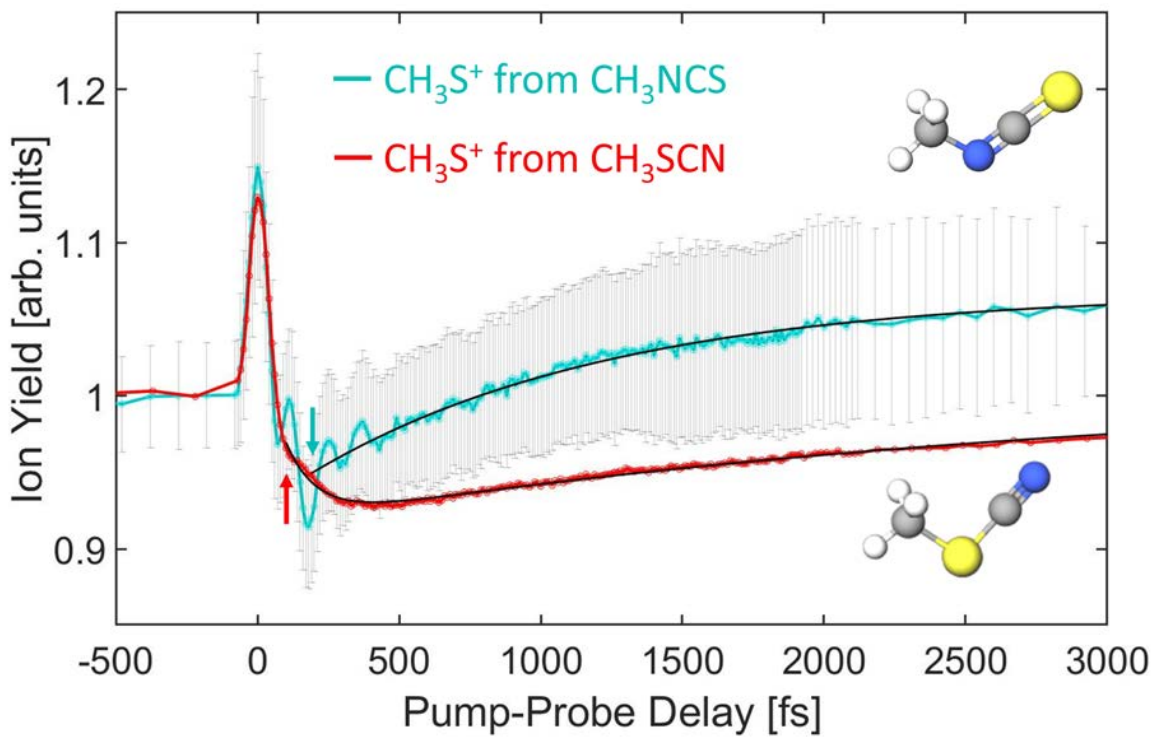

**Fig. 3**. The time-resolved yield of $CH_3S^+$ from $CH_3NCS$ (blue) and $CH_3SCN$ (red) following ultrafast ionization. Both ion yield traces are normalized such that the yield at long negative time delays is unity. The yield of $CH_3S^+$ from $CH_3NCS$ has been fit to a single exponential from the minima of the decay to the end of the trace; this is shown as a black line overlaying the blue curve. Error bars for the $CH_3S^+$ yield from $CH_3NCS$ show ±1 standard deviation. The yield of $CH_3S^+$ from $CH_3SCN$ has been fit to a biexponential from 100 fs to the end of the trace; this is shown as a black line overlaying the red curve. The starting points for the two fitted curves are indicated with red and blue arrows.

Fitting the transient associated with $CH_3S^+$ formation from MITC shown in Fig. 3, we obtain three exponential terms, a 35 ± 2 fs decay, a 1141 ± 62 fs rise, and a very long rise that causes the difference in ion yield between long negative and long positive times. The fitting parameters are summarized in Table II for MITC and MTC. The formation of $CH_3S^+$ from MTC is quite different, suggesting that the results obtained from MITC are the result of an intramolecular rearrangement and not an impurity in the sample by the isomer. The production of $CH_3S^+$ from MTC proceeds with significantly slower dynamics, starting with a 50 ± 1 fs decay and a 3640 ± 230 fs rise, and followed by a very long rise, again corresponding to a minor difference in ion yield and negative and positive time delays.

**Table II.** Fitting parameters based on Eqs. (1) and (2) with coherent vibration frequencies, $v_1$ and $v_2$, for MITC ≡ $CH_3NCS$. $\tau_{decay}$ is the $\tau$ from Eq. (1) that quantifies the timescale of ion yield decay after time zero. $\tau_{rise1}$ is the $\tau$ from Eq. (1) that quantifies the timescale of ion yield recovery following the decay. $v_1$ and $v_2$ quantify the coherent oscillations found in the yields of almost all ions in the $CH_3NCS$ mass spectrum. The angular frequencies of the oscillations have been converted to wavenumbers. No oscillations were observed in MTC ≡ $CH_3SCN$.

| Parameter | $CH_3NCS$ | $CH_3SCN$ |
|---|---|---|
| $\tau_{decay}$ (fs) | 35 ± 2 | 50 ± 1 |
| $\tau_{rise1}$ (fs) | 1141 ± 62 | 3640 ± 230 |
| $v_1$ (cm$^{-1}$) | 123 | — |
| $v_2$ (cm$^{-1}$) | 290 | — |

The oscillatory component observed for MITC (see the blue line in Fig. 3) was isolated from the time-dependent yields of the product ion ($CH_3S^+$) and the molecular ion ($CH_3NCS^+$) by subtracting the slow exponential components from the observed experimental data. The residuals are shown in Fig. 4 (see Fig. 4 (a)). Residual traces for several other ions are shown in Fig. S1 in the Supplementary Material. We find that the product and molecular ion oscillations are out of phase. Fourier analysis of the residuals allows us to identify a range of possible frequencies for the residual oscillations. Zero padding is applied to the experimental data before the fast Fourier transform to ensure an appropriate number of data points, but since no new information is being added, the spectrum remains broad. To refine the data analysis, a technique called the maximum entropy method (MEM)[50] is used to extract the sample frequencies from the broad Fourier transform. The MEM approach extracts two beating frequencies at ~123 cm$^{-1}$ and ~290 cm$^{-1}$, shown in Fig. 4 (b).

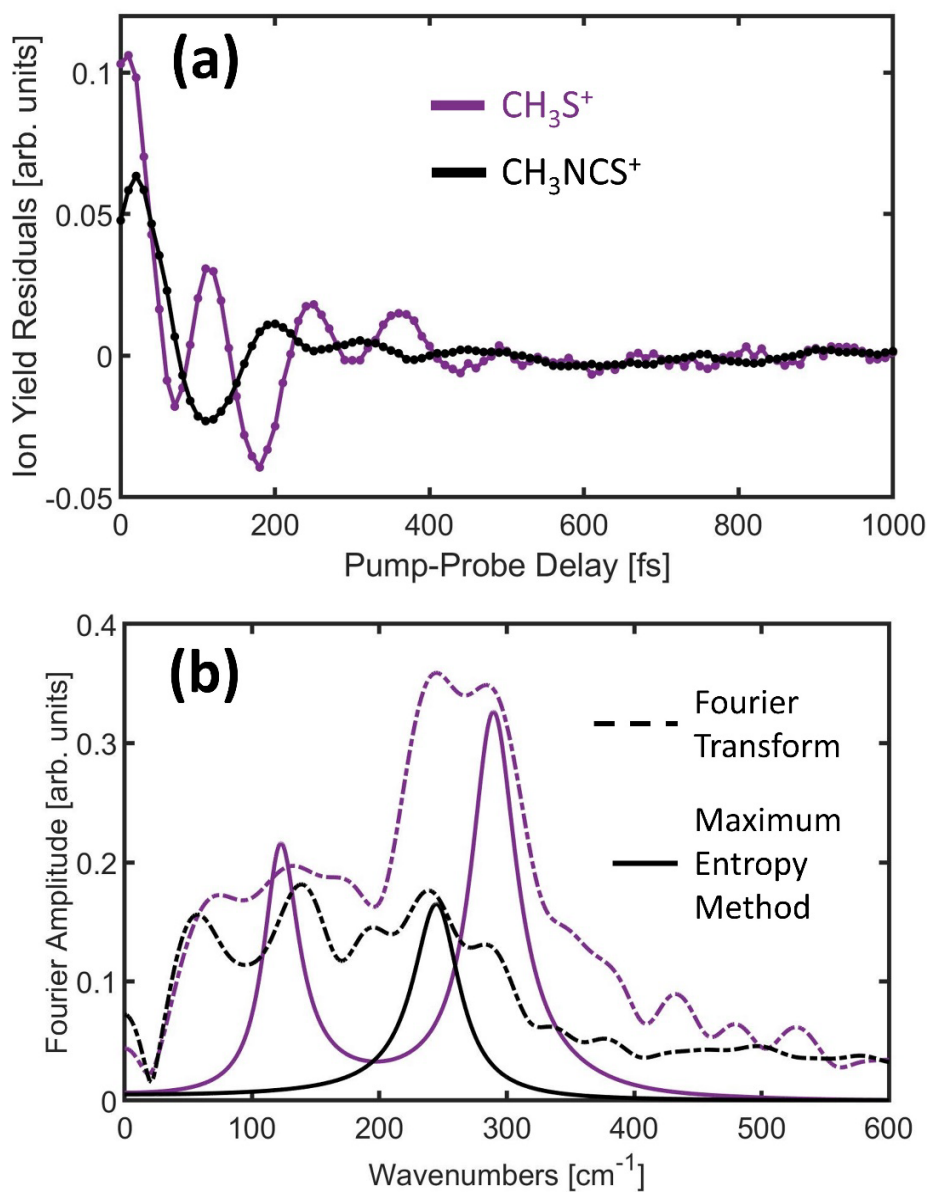

**Fig. 4**. Coherent vibrational motion observed in the molecular ion for $CH_3NCS^+$ (black) and the rearrangement reaction product $CH_3S^+$ (purple). (a) Residuals after fitting $CH_3NCS^+$ and $CH_3S^+$ yields. (b) Fast Fourier transform and maximum entropy analysis of the residuals shown in (a), with zero padding to 2,048 data points.

We find that the rearrangement reaction in the $CH_3NCS^+$ ion takes place in $1.14 \pm 0.06$ ps, while the bond cleavage in the $CH_3SCN^+$ ion takes place in $3.64 \pm 0.23$ ps. These differences in timescale corroborate the idea that $CH_3S^+$ generation from MITC is not due to the contamination from its MTC isomer, because if the existence of $CH_3S^+$ were from the MTC impurity, then one would expect similar timescales for the rearrangement reactions. Moreover, the difference in timescales contradicts intuition, as one would expect the MTC to undergo the reaction quicker, as it only involves a single bond breaking event, whereas MITC requires a significant intramolecular rearrangement prior to the S–CN bond breaking. As mentioned above, the generation of $CH_3S^+$ from MITC shows coherent oscillations which, after analysis, reveal two beating frequencies at ~123 cm$^{-1}$ and ~290 cm$^{-1}$. When analyzing the oscillations observed in the parent $CH_3NCS^+$ ion, only one oscillation frequency, at ~250 cm$^{-1}$, is retrieved (see the black solid line in Fig. 4 (b)). This is due to the weaker oscillation amplitude in the $CH_3NCS^+$ signal that likely results from the multiple fragmentation pathways available to this molecular ion. It is likely that the single oscillation frequency is an average of multiple beating frequencies that cannot be resolved by Fourier transform or by MEM.

To better understand why the two frequencies are observed in the $CH_3S^+$ product ion traces, we examined the harmonic frequencies that correspond to the C–N–C bending modes in the $CH_nNCS^+$ species obtained in our B3LYP/cc-pVTZ computations (see Fig. 5). Our calculations reveal the existence of two low-frequency C–N–C bending modes for the $CH_3NCS^+$ ion that closely match the beating frequencies found through the experimental measurement of the time-dependent ion yields and the subsequent Fourier analysis. The C–N–C bending modes in the $CH_3NCS^+$ ion, shown in Fig. 5, correlate well with the intramolecular rearrangements that the ion would need to undergo to modulate the experimentally observed yield of $CH_3S^+$. Furthermore, while the ground state of the parent $CH_3NCS$ species is bent, the corresponding ion state has a linear C–N–C–S backbone in its ground-state geometry. As a result, the geometric relaxation of $CH_3NCS^+$ after the vertical ionization event matches the motion of atoms characterizing the C–N–C bending modes in this species. All of this suggests that the two beating frequencies at ~123 cm$^{-1}$ and ~290 cm$^{-1}$ extracted from our experiments are remnants of the motions in $CH_3NCS^+$ that promote the rearrangements of atoms resulting in the formation of the $[CH_3S \cdots CN]^+$ intermediate. Interestingly, the remaining $CH_nNCS^+$ ions with $n = 0–2$ exhibit the analogous C–N–C bending motions in the 100–200 cm$^{-1}$ frequency range as well, which suggests that the $CH_nS^+$ fragments may also be formed from the $CH_nNCS^+$ species with $n = 0–2$.

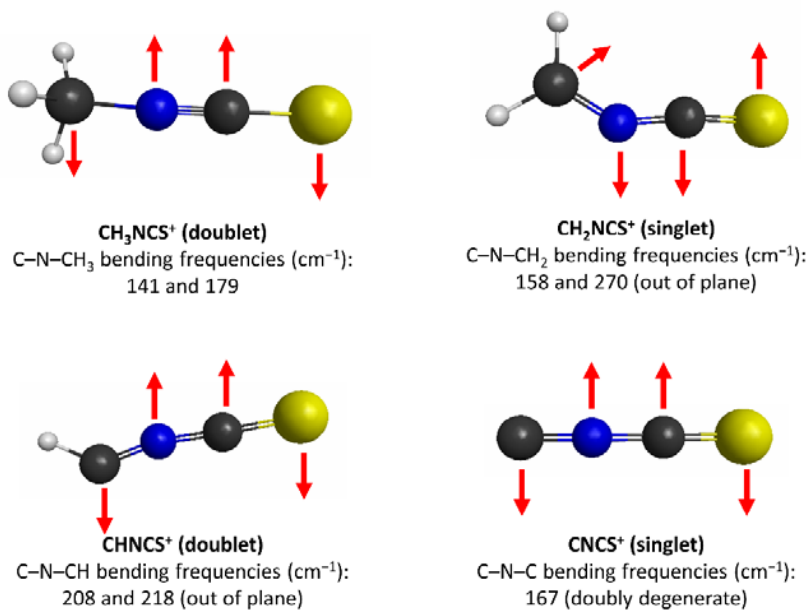

**Fig. 5.** Visual representations of the optimized geometries and normal modes corresponding to the C–N–C bending in the $CH_3NCS^+$, $CH_2NCS^+$, $CHNCS^+$, and $CNCS^+$ species, along with the associated harmonic frequencies obtained in the B3LYP/cc-pVTZ computations.

The above observations encouraged us to look for possible intramolecular rearrangement pathways of the $CH_nNCS^+$ ions with $n = 0$–$3$ that correlate with the C–N–C bending motions and that might lead to the formation of the $CH_nS^+ + CN$ products. As explained in Section II, we did so by performing IRC scans for each of the $CH_nNCS^+$ species. The resulting rearrangement pathways are shown in Figs. 6 ($CH_3NCS^+$) and 7 (all $CH_nNCS^+$ species). As shown in Fig. 6, the IRC pathway associated with the C–N–C bending motion in $CH_3NCS^+$ may result in the formation of the $[CH_3S \cdots CN]^+$ intermediate, which could then produce $CH_3S^+ + CN$ by breaking the S–CN bond. This could explain the observed production of $CH_3S^+$ following strong-field ionization of $CH_3NCS$. The question remains how the $CH_3NCS^+$ ion can overcome the ~3.5 eV barrier characterizing the $CH_3NCS^+ \to [CH_3S \cdots CN]^+$ rearrangement process and the additional energy necessary to subsequently break the S–CN bond to form $CH_3S^+$ and CN. While the answer to this question requires further studies, we may hypothesize that electron rescattering results in sufficient internal energy to populate excited states of the ionized species, which could relax non-radiatively to the lowest state of the $CH_3NCS^+$ ion and drive the reaction forward. As shown in Fig. 6, our IP-EOMCC(3h-2p) calculations indicate the existence of several excited states of $CH_3NCS^+$ that are energetic enough to overcome the reaction barrier and reach the $CH_3S^+ + CN$ products.

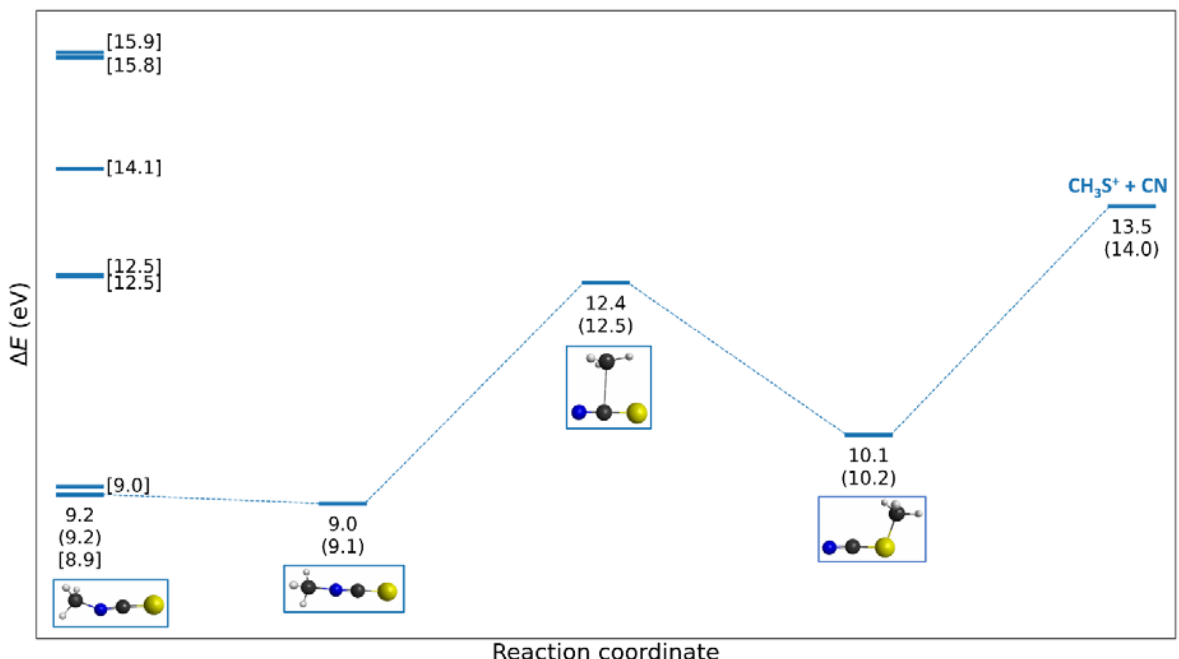

**Fig. 6**. The intramolecular rearrangement pathway leading to the formation of the $CH_3S^+$ + CN products from the $CH_3NCS^+$ species, along with the vertical ionization energies characterizing $CH_3NCS$. The energies along the $CH_3NCS^+ \rightarrow CH_3S^+$ + CN reaction pathway were obtained with B3LYP and CR-CC(2,3) (numbers in parentheses). The vertical ionization energies characterizing $CH_3NCS$ (numbers in square brackets) were obtained using IP-EOMCC(3h-2p). All energies are reported relative to the neutral $CH_3NCS$ molecule in its ground electronic state.

The intramolecular rearrangement pathways that result in the formation of the $CH_nS^+$ + CN products from the $CH_nNCS^+$ species with $n = 0$–3 are shown together in Fig. 7. The energies of the stationary points shown in the figure are the CR-CC(2,3) values relative to the respective minima on the $CH_nNCS^+$ potential energy surfaces. Assuming that the $CH_3NCS^+$ species loses hydrogen(s) prior to the intramolecular rearrangement and that one is able to access excited states of the $CH_nNCS^+$ ions with enough energy to overcome the relevant reaction barriers, the formation of the $CH_nS^+$ products with $n = 0$–3 is possible. It is worth pointing out that the relative energies of the final $CH_nS^+$ + CN products correlate well with the ion yields observed in Fig. 1, with $CHS^+$ being most easily formed, followed by $CH_2S^+$ and $CH_3S^+$. The fact that $CS^+$, being the most difficult to form, does not correspond to its experimentally observed ion yield relative to the remaining $CH_nS^+$ products with $n = 1$–3 indicates that its formation may involve other reaction pathways which we have not identified in our computations. We plan to return to this issue in our future investigations.

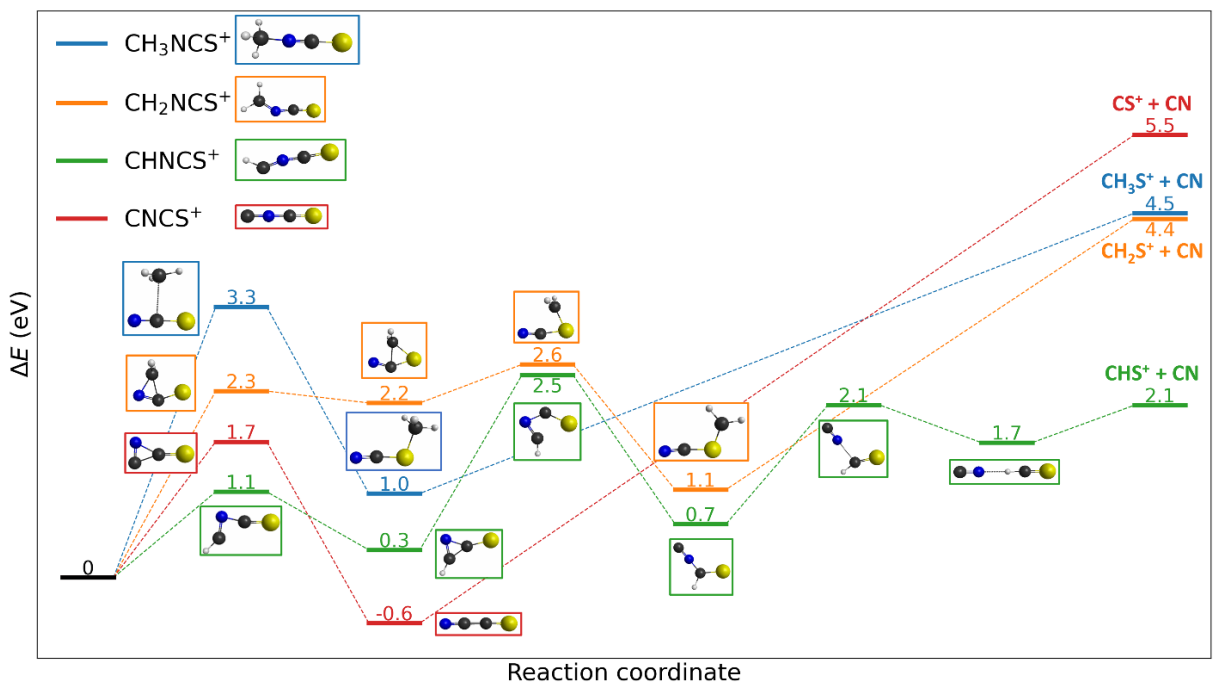

**Fig. 7**. The intramolecular rearrangement pathways leading to the formation of the CH$_n$S$^+$ species with $n$ = 0–3 from the corresponding CH$_n$NCS$^+$ ions. For each pathway, the reported energies are the CR-CC(2,3) values relative to the CH$_n$NCS$^+$ parent ion.

## IV. Conclusions

We demonstrated that the CH$_3$NCS (MITC) molecule can generate the CH$_3$S$^+$ species upon strong-field tunnel ionization. This implies an intramolecular rearrangement in the CH$_3$NCS$^+$ ion state, which warrants further analysis. The yield of the CH$_3$S$^+$ product ion was tracked by using disruptive probing following strong-field ionization, resulting in decreased yield after time zero that recovers on a picosecond timescale. The dynamics of the CH$_3$NCS$^+$ ion were found to be very different than those of its CH$_3$SCN$^+$ isomer, refuting the possibility of the formation of CH$_3$S$^+$ as a result of CH$_3$SCN contamination. The fact that the formation of CH$_3$S$^+$ from CH$_3$NCS$^+$ through complex intramolecular rearrangement discussed in the present study is faster than that in CH$_3$SCN$^+$, which involves a rather simple single bond cleavage, defies intuition and highlights the importance and usefulness of time-resolved studies of reaction dynamics involving ionic radical species. We will return to the examination of the formation of CH$_n$S$^+$ ($n$ = 0–3) species from the CH$_3$NCS$^+$ parent ion, with the hope of obtaining further insights, in our future studies. The yield of CH$_3$S$^+$ following strong-field ionization of CH$_3$NCS exhibited coherent oscillations which, after Fourier analysis, revealed two beating frequencies that correspond to the C–N–C bending modes in the ion state. This was corroborated by quantum chemical computations, which also helped us determine possible intramolecular rearrangement mechanisms of the $CH_nNCS^+$ species with $n = 0–3$ that might result in the formation of the $CH_nS^+$ ions after strong-field ionization of $CH_3NCS$. The results reported in this work provide a specific example of the type of information that disruptive probing provides about the timescale of product formation following high-energy (>15 eV) excitation via strong-field ionization or 70 eV electron ionization. This method could potentially be used to elucidate ion fragment formation mechanisms in mass spectrometry.

**Supplementary Material**

Residual oscillations in the yields of several key ions seen in the mass spectra of $CH_3NCS$ obtained in this work and the nuclear coordinates of the stationary points along the $CH_nNCS^+ \rightarrow CH_nS^+ + CN$ pathways obtained in the B3LYP/cc-pVTZ optimizations.

**Authors' Contributions**

JS, SL, BJ, and MD carried out the experiments, whereas SHY, SSP, and PP were responsible for the computations. SL suggested to study $CH_3NCS$ and acquired the preliminary experimental data, JS and BJ carried out all the reported experiments under the supervision of MD, and SHY and SSP performed all the reported quantum chemistry computations under the supervision of PP. All authors, except SL, contributed to the interpretation of results and writing of the manuscript.

**Acknowledgments**

The experimental part of this study is based upon work supported by the U.S. Department of Energy, Office of Science, Office of Basic Energy Sciences, Atomic, Molecular and Optical Sciences Program under Award Number SISGR (DE-SC0002325) to MD. The computational work was supported by the Chemical Sciences, Geosciences and Biosciences Division, Office of Basic Energy Sciences, Office of Science, U.S. Department of Energy (Grant No. DE-FG02-01ER15228 to PP). The authors thank Drs. Ilias Magoulas and Jun Shen for their help in the design of the computational protocol in the early discussions of the project.

**Data Availability**

The data that support the findings of this study are available within the article and its Supplementary Material. Further data are available from the corresponding author upon reasonable request.

# Supplementary Material for

# Femtosecond intramolecular rearrangement of the CH$_3$NCS radical cation

Jacob Stamm[1], Shuai Li[1], Bethany Jochim[1], Stephen H. Yuwono[1], Swati S. Priyadarsini[1], Piotr Piecuch[1,2], and Marcos Dantus[1,2,*]

[1] Department of Chemistry, Michigan State University, East Lansing, MI 48824, USA
[2] Department of Physics and Astronomy, Michigan State University, East Lansing, MI 48824, USA
* Corresponding author: MD email: dantus@msu.edu

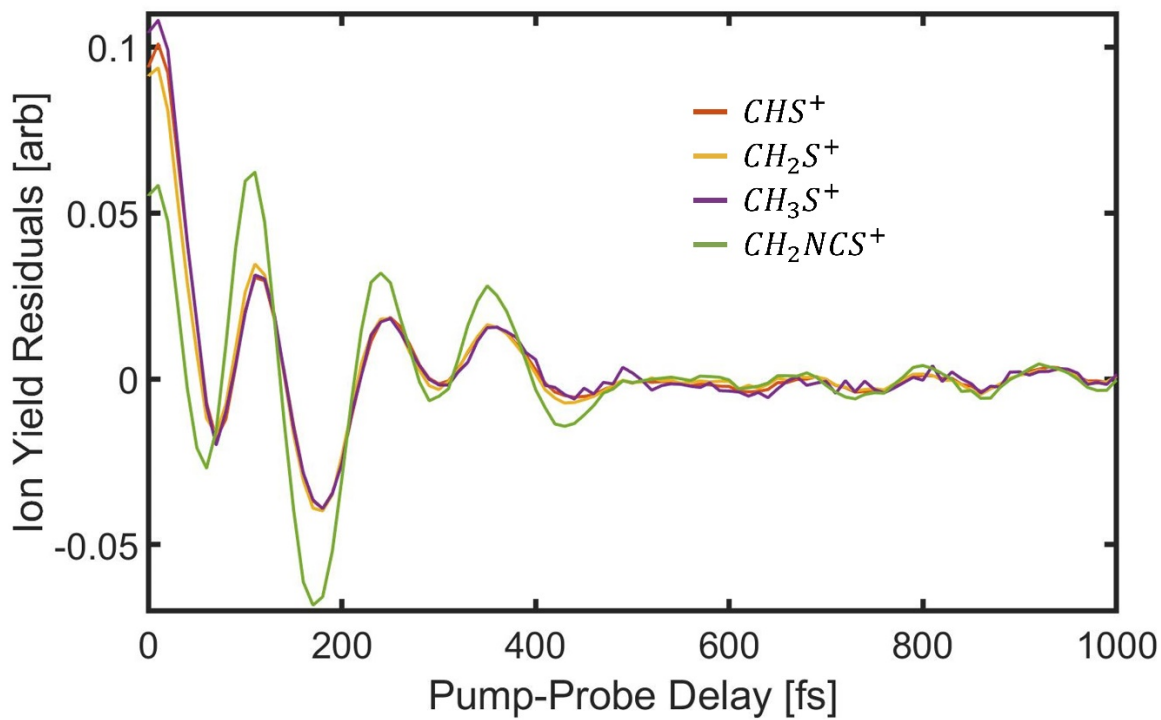

**Fig. S1**. Residual ion yields for several ions with extra hydrogen loss relative to the $CH_3NCS^+$ and $CH_3S^+$ ions analyzed in the paper. Unlike $CH_3NCS^+$ (see Fig. 4 (a) of the main text), the $CH_2NCS^+$ species shows strong in-phase oscillations with the product $CH_nS^+$ ions with $n = 1–3$.

The Cartesian coordinates (in Å) defining the geometry of the CH$_3$NCS molecule and the stationary points along each of the CH$_n$NCS$^+$ → CH$_n$S$^+$ + CN ($n$ = 0–3) reaction pathways obtained in B3LYP/cc-pVTZ optimizations.

**CH$_3$NCS (singlet)**

```
       X              Y              Z
C  -0.6465859421   0.0728571310  -0.0176069169
C   1.7863556443   0.3134905181  -0.6825813532
N   0.6282019205   0.0723347533  -0.6507458028
S   3.3389649199   0.5951682540  -0.8180359342
H  -0.5864051737  -0.4083086908   0.9605953262
H  -1.3641822124  -0.4706641554  -0.6301685466
H  -1.0048809865   1.0957175900   0.1129508876
```

**CH$_3$NCS$^+$ (doublet) - Reactant**

```
       X              Y              Z
C  -1.3071362494   1.7703688307   0.7812792420
C   0.0748859797  -0.2826022203   0.0550717201
N  -0.5435903393   0.6422011220   0.3821505980
S   0.9414846651  -1.5769722092  -0.4017418075
H  -0.6334884907   2.5464403320   1.1448625980
H  -1.9612868896   1.4505368205   1.5984272205
H  -1.9090178656   2.1152867159  -0.0582861484
```

**CH$_3$NCS$^+$ - Transition state**

```
       X              Y              Z
C  -1.4844258014   0.9262545216  -0.3911973444
C   0.4224937566  -0.1179145253   0.6529904979
N   0.3916259115   0.6679710883   1.5305938153
S   0.4720691966  -1.2319609171  -0.5658566802
H  -1.0098962604   1.8822335988  -0.2089726999
H  -2.1160413572   0.4967668278   0.3776022222
H  -1.5083674804   0.5199432237  -1.3929800174
```

**CH$_3$NCS$^+$ - Intermediate**

```
       X              Y              Z
C  -1.3017945375   0.1699437847  -1.0547426755
C   0.5582453257  -0.0548372593   0.9064541631
N   0.9261272673   0.4169713193   1.9089376156
S   0.1199540933  -0.8028672201  -0.5064607399
H  -0.9783935561   1.2006888099  -1.2184786373
H  -2.0746988696   0.1288404717  -0.2839593962
H  -1.6282497756  -0.3013913038  -1.9802468865
```

**CH₂NCS⁺ (singlet) - Reactant**

|   | X | Y | Z |
|---|---|---|---|
| C | 1.6408186739 | -1.7661509483 | 0.5633505271 |
| C | 0.3605679871 | 0.1238084171 | -0.1812819878 |
| N | 0.6696315981 | -1.0175245101 | 0.2139328103 |
| S | -0.1931995506 | 1.4508102010 | -0.6632854470 |
| H | 1.4269671009 | -2.7886969872 | 0.8602646161 |
| H | 2.6670041806 | -1.4007201005 | 0.5684648916 |

**CH₂NCS⁺ - Transition state 1**

|   | X | Y | Z |
|---|---|---|---|
| C | 1.4718394134 | -0.5502030984 | 0.0771157890 |
| C | -0.1525025706 | -0.2569583552 | -0.0127641657 |
| N | 0.2055476042 | -1.3442966779 | 0.4112181353 |
| S | 0.2745512325 | 1.1790594421 | -0.5877986523 |
| H | 1.9636148336 | -0.8682659683 | -0.8336538570 |
| H | 1.9929531487 | -0.1655150757 | 0.9450536034 |

**CH₂NCS⁺ - Intermediate 1**

|   | X | Y | Z |
|---|---|---|---|
| C | 1.4798025697 | -0.2477123541 | -0.0429072123 |
| C | -0.1953944792 | -0.3357009078 | 0.0185917767 |
| N | 0.0810918976 | -1.4173977955 | 0.4423979021 |
| S | 0.3426346875 | 1.1218869199 | -0.5661042933 |
| H | 1.9392786680 | -0.7981159065 | -0.8534232367 |
| H | 1.9664512982 | -0.1007099800 | 0.9122214522 |

**CH₂NCS⁺ - Transition state 2**

|   | X | Y | Z |
|---|---|---|---|
| C | -1.2037505091 | 0.5981436911 | -0.2220307216 |
| C | 0.4505059805 | 0.0774646139 | 0.7421628441 |
| N | 0.6664926247 | 0.6395181624 | 1.7454194438 |
| S | -0.1421247649 | -0.7056739886 | -0.6448057071 |
| H | -0.8705031278 | 1.6263017988 | -0.3194727339 |
| H | -2.2247947577 | 0.3875963165 | 0.0938866855 |

**CH₂NCS⁺ - Intermediate 2**

|   | X | Y | Z |
|---|---|---|---|
| C | -1.5159256007 | 0.3865935017 | -0.7329808266 |
| C | 0.5142839337 | 0.1070317276 | 0.8935519197 |
| N | 1.0841932028 | 0.4678255388 | 1.8309572129 |
| S | -0.2058552367 | -0.5412222052 | -0.5014630955 |
| H | -1.7883564242 | 1.2142402863 | -0.0866925749 |
| H | -2.1312211384 | 0.1350154619 | -1.5935299706 |

**CHNCS⁺ (doublet) - Reactant**

|   | X | Y | Z |
|---|---|---|---|
| C | -2.3011129671 | -0.5463411689 | -0.0302794785 |
| C |  0.0602663798 |  0.0717636758 |  0.0038578212 |
| N | -1.1729140945 | -0.1029166971 | -0.0059673784 |
| S |  1.5513000953 |  0.3531843194 |  0.0195336532 |
| H | -3.3003773117 | -0.1151930252 | -0.0072979949 |

**CHNCS⁺ - Transition state 1**

|   | X | Y | Z |
|---|---|---|---|
| C | -1.5821903938 | -0.7504679765 | -0.0410162542 |
| C | -0.0976331627 |  0.2030961703 |  0.0109397643 |
| N | -1.4185811122 |  0.4938799884 |  0.0260276678 |
| S |  1.4181686143 |  0.1639904824 |  0.0092935606 |
| H | -2.3681553945 | -1.5064757901 | -0.0817093618 |

**CHNCS⁺ - Intermediate 1**

|   | X | Y | Z |
|---|---|---|---|
| C | -1.4130390078 | -0.6783164880 | -0.0370215564 |
| C | -0.1843640546 | -0.0178981764 | -0.0009906558 |
| N | -1.4549375476 |  0.5973553292 |  0.0315682289 |
| S |  1.3952855125 |  0.1824561879 |  0.0102639728 |
| H | -2.0209064247 | -1.5756070046 | -0.0854188047 |

**CHNCS⁺ - Transition state 2**

|   | X | Y | Z |
|---|---|---|---|
| C | -1.6171739911 | -0.0467847284 |  0.0084600628 |
| C | -1.1218608776 |  1.5635226641 | -0.1821120953 |
| N | -2.3363214649 |  0.9549582523 |  0.2079261081 |
| S |  0.3369066837 |  0.8616975523 |  0.0230185504 |
| H | -1.6173708900 | -1.0867862004 | -0.3149724660 |

**CHNCS⁺ - Intermediate 2**

|   | X | Y | Z |
|---|---|---|---|
| C | -1.0843765380 |  0.2519691807 |  0.0764687820 |
| C | -2.7185632644 |  2.1457271869 | -0.1885664344 |
| N | -1.9173824262 |  1.2681713303 | -0.0645956218 |
| S |  0.5513272376 |  0.3794113163 |  0.1040761925 |
| H | -1.5124767017 | -0.7507448782 |  0.1831270933 |

**CHNCS⁺ - Transition state 3**

|   | X | Y | Z |
|---|---|---|---|
| C | -0.8084356874 | 2.5235195734 | 0.0755909795 |
| C | -2.8040601103 | 4.9522290930 | -0.2650943790 |
| S | 0.6359887484 | 2.8785317083 | 0.2965630361 |
| N | -2.0658394536 | 4.0699013510 | -0.1391324190 |
| H | -1.6261369146 | 1.8311000681 | -0.0435315179 |

**CHNCS⁺ - Intermediate 3**

|   | X | Y | Z |
|---|---|---|---|
| C | -0.2263036115 | 3.0689235252 | 0.1600430914 |
| C | -3.7217161907 | 5.1285976466 | -0.4102128070 |
| S | 1.0344389037 | 2.3281078735 | 0.3657424614 |
| N | -2.7385338460 | 4.5467060689 | -0.2496986611 |
| H | -1.1755641143 | 3.6268028177 | 0.0051917148 |

**CNCS⁺ (singlet) - Reactant**

|   | X | Y | Z |
|---|---|---|---|
| C | -0.0971189462 | -0.2269240065 | -0.0779401030 |
| C | 0.1011884593 | 1.3213149471 | 1.8362579092 |
| S | -0.2179078057 | -1.1638782374 | -1.2384435112 |
| N | 0.0047052006 | 0.5634399549 | 0.9008212121 |

**CNCS⁺ - Transition state**

|   | X | Y | Z |
|---|---|---|---|
| C | -0.1093506271 | -0.0764477865 | 0.0236880675 |
| C | -0.1696649178 | 1.1286013882 | 0.9404391074 |
| S | -0.2642154992 | -0.9634738723 | -1.1829103997 |
| N | 0.3449941075 | 0.1556704381 | 1.4453135133 |

**CNCS⁺ - Intermediate**

|   | X | Y | Z |
|---|---|---|---|
| C | -0.1103043013 | -0.2518615060 | -0.1354815914 |
| C | 0.1244482872 | 0.3670005868 | 1.0450754410 |
| S | -0.3683608394 | -0.9322101701 | -1.4332279615 |
| N | 0.3260682527 | 0.8982657549 | 2.0587722185 |

**CN (doublet)**

|   | X | Y | Z |
|---|---|---|---|
| N | -1.8351246572 | 3.7589767027 | -0.0076560300 |
| C | -2.7275735342 | 4.4673754684 | -0.2379587858 |

**CS⁺ (doublet)**

|   | X | Y | Z |
|---|---|---|---|
| C | -1.0470183427 | 0.3567979867 | -0.1027586960 |
| S | 0.4315332065 | 0.2244686173 | -0.0709965058 |

**CHS$^+$ (singlet)**

|   | X | Y | Z |
|---|---|---|---|
| C | -0.7971021058 | -0.0993189021 | 0.1258992680 |
| S | 0.5098693845 | 0.5757325978 | 0.0719739336 |
| H | -1.7580465689 | -0.5957992414 | 0.1658117590 |

**CH$_2$S$^+$ (doublet)**

|   | X | Y | Z |
|---|---|---|---|
| C | 1.0034589935 | 0.9262603843 | 0.1305273056 |
| S | 1.8531298787 | 0.3331647839 | 1.3328405200 |
| H | 1.4738103261 | 1.1783238141 | -0.8250096967 |
| H | -0.0730663889 | 1.0979851578 | 0.2278953297 |

**CH$_3$S$^+$ (triplet)**

|   | X | Y | Z |
|---|---|---|---|
| C | -1.0222827066 | 1.8332245118 | -1.0660071530 |
| S | 0.3064037799 | 1.0011905247 | -0.3128064754 |
| H | -0.6986471872 | 2.8444253665 | -1.3519877195 |
| H | -1.8571542018 | 1.8964472412 | -0.3531768719 |
| H | -1.3366776051 | 1.2751186697 | -1.9598346371 |